\documentclass[preprint]{elsarticle}
\usepackage{hyperref}

\journal{}

\usepackage{paralist}
\usepackage{graphicx}
\usepackage{csquotes}
\usepackage{paralist}
\usepackage{microtype}
\usepackage{siunitx}
\usepackage{url}
\usepackage{multirow}
\usepackage{booktabs}

\usepackage{lscape}
\usepackage{rotating}
\usepackage{listings}
\usepackage{lstautogobble}
\usepackage{pythonhighlight}
\usepackage{todonotes}
\usepackage{tabularx}
\usepackage{subcaption}
\usepackage{amsmath}
\usepackage{changes}

\newcommand{\stage}[1]{\emph{#1}-Stage}

\newcommand\YAMLkeystyle{\color{black}\bfseries}

\lstdefinelanguage{yaml}{
	basicstyle={\scriptsize\ttfamily},
	comment=[l]{\#},
	commentstyle=\color{gray}\ttfamily,
	morestring=[b]',
	morestring=[b]"
}

\newcommand\TimeForYAML{\lst@AddToHook{EOL}{\YAMLkeystyle}}

\lstdefinelanguage{Python}{
	basicstyle={\scriptsize\ttfamily},
	comment=[l]{\#},
	commentstyle=\color{gray}\ttfamily,
	morestring=[b]',
	morestring=[b]"
}

    \makeatletter
\def\ps@pprintTitle{%
	\let\@oddhead\@empty
	\let\@evenhead\@empty
	\def\@oddfoot{\reset@font\hfil\thepage\hfil}
	\let\@evenfoot\@oddfoot
}
\makeatother

\begin{document}
	\begin{frontmatter}
		\title{An Artifact-based Workflow for Finite-Element Simulation Studies}
		
		\author[mosiaddress]{Andreas Ruscheinski\corref{cor1}}
		\ead{andreas.ruscheinski@uni-rostock.de}
		\author[mosiaddress]{Pia Wilsdorf}
		
		\author[etecaddress]{Julius Zimmermann}
		\author[etecaddress]{Ursula van Rienen}
		\author[mosiaddress]{Adelinde M. Uhrmacher}
		
		\address[mosiaddress]{Institute for Visual and Analytic Computing, University of Rostock, 18059 Germany}
		\address[etecaddress]{Institute of General Electrical Engineering, University of Rostock, 18059 Germany}
		
		\cortext[cor1]{Corresponding author}
		\begin{abstract}
			Workflow support typically focuses on single simulation experiments. This is also the case for simulation based on finite element methods.
			If entire simulation studies shall be supported, flexible means for intertwining revising the model, collecting data, executing and analyzing experiments are required.
			Artifact-based workflows present one means to support entire simulation studies, as has been shown for stochastic discrete-event simulation.
			To adapt the approach to finite element methods, the set of artifacts, i.e., conceptual model, requirement, simulation model, and simulation experiment, and the constraints that apply are extended by new artifacts, such as geometrical model, input data, and simulation data.
			Artifacts, their life cycles, and constraints are revisited revealing features both types of simulation studies share and those they vary in.
			Also, the potential benefits of exploiting an artifact-based workflow approach are shown based on a concrete simulation study.
			To those benefits belong guidance to systematically conduct simulation studies, reduction of effort by automatically executing specific steps, e.g., generating and executing convergence tests, and support for the automatic reporting of provenance.
		\end{abstract}
		
		\begin{keyword}
			Artifact-based Workflow\sep Finite Element Analysis\sep Simulation Studies\sep Provenance\sep Knowledge-based engineering\sep Process Engineering
		\end{keyword}
	\end{frontmatter}
	
	\section{Introduction}
	\label{sec:introduction}
	Modeling and simulation, including finite element analysis (FEA), has become an established method for engineering and research.
	Typically, larger research and engineering projects require to conduct and intertwine various simulation studies (\emph{in silico}), with their own research question, with laboratory experiments (\emph{in vitro}) acquiring the data needed to validate the simulation results or to test research hypotheses.
	Thus, the results of the simulation studies and laboratory experiments have to be put into relation to each other.
	For example, in the context of electrically active implants \emph{in vitro} and \emph{in silico} studies are combined to study the impact of electric fields on cellular dynamics.
	In \cite{Krueger2019}, the authors first conduct a simulation study including various simulation experiments that reveal the distribution and strengths of the electrical field (EF).
	Afterwards, the cellular responses that are observed in \emph{in vitro} experiments are related to the distribution of the EF.
	The results can be used to subsequently conduct more detailed simulation studies, e.g., to study the EF's impact on membrane related dynamics, intracellular dynamics, and cellular function \cite{thrivikraman2018unraveling,farooqi2019numerical}.
	
	The intricate nature of simulation studies, interleaving model refinement, the execution of simulation experiments, and the acquisition of data, leads to the question of how to support conducting and documenting these studies~\cite{fujimoto2017research}.
	Answers rely on a systematic specification of the process in terms of its resources, products, and activity patterns, i.e., a workflow.
	
	So far, workflows have been successfully exploited to support single
	computational tasks within simulation studies, e.g.,
	to execute simulation experiments and to analyze simulation results~\cite{gorlach2011conventional, Sonntag2013, Rybacki2011, Ribault2015,chopard2014framework}.
	However, entire simulation studies intertwine highly interactive processes, e.g., objective and requirements specification, knowledge acquisition, data collection, and model development, with computational processes, e.g., experiment execution, and data analysis, where each decision in each process may influence the study outcomes \cite{Balci2012}.
	Thus, more flexible workflows are required to support entire simulation studies.
	
	In \cite{ruscheinski2019artifact}, a declarative, artifact-based workflow approach achieves the required flexibility to capture the intricate nature of stochastic discrete-event simulation studies while adhering to the constraints that apply between the different activities and to the relations between the different work products, i.e., the conceptual model, requirements, simulation model, and simulation experiments.
	
	Here, we adapt the artifact-based workflow towards FEA simulation studies to explore how such a declarative workflow approach can contribute to the desired guidance, best practices, documentation of FEA simulation studies, reuse of results, and thus to the quality and credibility of a study's results~\cite{anderson2007verification, erdemir2012considerations, hicks2015my, cucurull2019best}.
	
	\section{An Artifact-based Workflow for FEA Simulation Studies}
	Artifact-based workflows are suitable for handling knowledge-intensive processes, such as simulation studies, where \enquote{a lot of knowledge and expertise is implicit, resides in experts' heads or it has a form of (informal) best practices or organizational guidelines}~\cite{di2015knowledge}.
	In artifact-based workflows, we describe processes as a set of artifacts interacting with each other.
	Therefore, each artifact consists of a lifecycle model and an information model \cite{hull2011business}.
	The lifecycle model describes how the artifact evolves in the process, whereas the information model structures the attributes and relations of the artifact.  
	
	The artifact-based Guard-Stage-Milestone meta-model (GSM) \cite{hull2011business} allows us to declaratively specify the lifecycles of artifacts \cite{vaculin2011declarative, DeGiacomo2015}.
	In the GSM, we represent the activities as \emph{stages} equipped with \emph{guards} and \emph{milestones}.
	The \emph{guards} specify the preconditions at which a stage can be entered to execute the activity. 
	The \emph{milestones} specify what is achieved by leaving a stage, i.e., the effect of the activity.
	During the execution of the process, we test the \emph{guards} and \emph{milestones} against the information models to determine which activities can be executed and what results are achieved.
	Moreover, the milestones are also stored as part of the information models of the artifacts to record what results are achieved.
	To use the GSM to specify an entire simulation study, we map the products of a simulation study \cite{fujimoto2017research,Balci2012},
	e.g., the conceptual model, the simulation models, and simulation experiments, to artifacts in our workflow.
	The lifecycles of these artifacts describe \emph{how} the products are developed and their information models \emph{what} is developed.
	This knowledge can then be used to implement support mechanisms for simulation studies, e.g., to guide the modeler through the simulation study, to document the provenance of the products and to generate simulation experiments (see Section \ref{sec:case-study}). 
	
	In the following, we revisit our artifact-based workflow for stochastic discrete-event simulation studies \cite{ruscheinski2019artifact} to summarize the existing artifacts and the constraints that apply between them.
	After that, we look at the specifics of FEA simulation studies and identify the required adaptions of the workflow. 
	Thereafter, we present the adapted workflow.
	
	\subsection{Revisiting the Artifact-based Workflow}
	In \cite{ruscheinski2019artifact}, we presented an artifact-based workflow for stochastic-discrete event simulation studies using the GSM \cite{hull2011business}.
	Here, we used the artifacts to represent the simulation model(s), the simulation experiment(s), and the conceptual model, which included a separate artifact for handling the requirements of the simulation study.
	
	\paragraph{The Lifecycle Models}
	In our workflow, we perceive the conceptual model as a container for all information that helps us to develop the simulation model \cite{Robinson2008b, Fujimoto2017, Wilsdorf2020}, e.g., real-world data the simulation model shall be validated with or information about the domain of interest.
	The lifecycle of the conceptual model artifact consists of four stages.
	Each simulation study starts at the \stage{Specifying objective}, where the modeler has to specify the overall objective of the simulation study, e.g., to create a validated simulation model or to test hypotheses on a simulation model.
	After specifying the objective, the stage reaches its milestones which allows the modeler to continue by assembling the conceptual model or to create a new simulation model artifact in the \stage{Creating simulation model}.
	In the \stage{Assembling conceptual model}, the modeler can create a new requirement artifact or fill the conceptual model with information.
	The modeler can validate the conceptual model in the \stage{Validating conceptual model}.
	Therefore, we require that the conceptual model and its requirements have to be assembled, indicated by the milestones of the corresponding assembling-stages.
	During the validation, the modeler checks the consistency of the collected information and the requirements.
	After this, the stage reaches a milestone according to the validation result.
	To ensure the consistency of the information in the conceptual model, we invalidate the milestones whenever the modeler changes the conceptual model in the \stage{Assembling conceptual model} or its requirements. 
	
	The requirement artifact specifies model behavior that can be checked by validation experiments or serve as the calibration target for the calibration experiments.
	By this, the requirement artifact forms a constituent of the conceptual model.
	In the workflow, the modeler specifies the requirement in the \stage{Assembling requirement}.
	In this stage, we require the modeler to specify the requirement type before the specification can be provided.
	For example, we can use a requirement to refer to a time series expressed as a temporal logic formula or to in-vitro data collected in a CSV-file.
	After finishing the specification of the requirement, the modeler can leave the assembling stage which achieves its milestone to indicate that the requirement has been assembled.
	
	The lifecycle of the simulation model artifact consists of four stages to capture the assembling of the simulation model, the creation of simulation experiments that should be executed with the simulation model, and the validation as well as calibration of the simulation model.
	The \stage{Assembling simulation model} encapsulates the process of specifying a simulation model.
	Therefore, the modeler has to specify the modeling formalism before the simulation model can be specified.
	For example, the modeler might choose ML-Rules \cite{Helms2017} for a rule-based model or NetLogo \cite{tisue2004netlogo} for an agent-based model.
	In the \stage{Choosing requirements}, the modeler can choose the requirements that apply for the simulation model from the requirements in the conceptual model.
	To leave the assembling stage, we require that the specification has to be syntactically correct according to the chosen formalism.
	The simulation experiments are created by entering the \stage{Creating simulation experiment}.
	We use the simulation experiments to calibrate or to validate the simulation model but also to generate simulation data or to analyze the model behavior for what-if scenarios.
	In the \stage{Calibrating simulation model} 
	calibration experiments associated with the simulation model are executed.
	Therefore, we require that the simulation model and the experiments have to be assembled.
	If a useful parameter assignment has been found, we leave the stage and the model is marked as successfully calibrated and the stage reaches the corresponding milestone.
	By entering the \stage{Validating simulation model}, the modeler triggers the execution of all validation experiments.
	Consequently, we require that all validation experiments and the simulation model are assembled but also that the conceptual model has to be validated.
	Possible outcomes of the calibrating- and validating-stage are success or failure.
	
	The lifecycle of the simulation experiment artifacts consists of two stages.
	In the \stage{Assembling simulation experiment} the modeler has to specify an approach for the experiment description, e.g., an experiment specification language such as SESSL \cite{Ewald2014} or a scripting language such as R \cite{thiele2010netlogo, thiele2014facilitating}.
	The modeler can also choose the role to indicate that the experiment forms for example a validation or calibration experiment.
	If this is the case, the modeler also has to select a requirement associated with the simulation model, e.g., time series the simulation model shall reproduce or specific spatio-temporal patterns that the simulation results shall exhibit \cite{RuscheinskiWSC2020}. 
	In a validation experiment, the requirement should be checked whereas in calibration experiments the requirement serves as the calibration target of the experiment.
	To leave the assembling stage, we require that the provided experiment specification has to be syntactically correct according to the specified experiment approach.
	Finally, the simulation experiments can be executed by entering the \stage{Executing experiment}.
	Consequently, the simulation experiment and the model have to be assembled.
	
	\paragraph{The Information Models}
	As the modeler moves through the different stages of the artifact lifecycles, the corresponding information models of the artifacts are filled with the information provided by the modeler and achieved milestones.
	Also, the relations between the different artifacts, e.g., the chosen requirements from the conceptual model in the simulation model artifacts, are stored as part of the information model.

	\subsection{Specifics of FEA simulation studies}
	Finite-element analysis simulation studies subsume all simulation studies using the finite-element method to simulate physical phenomena.
	Therefore, FEA simulation studies can be structured into three different steps \cite{roylance2001finite, datta2010introduction}: \begin{inparaenum} \item pre-processing, \item processing, and \item post-processing\end{inparaenum}.
	The pre-processing step comprises all activities required to specify the inputs in the FEA simulation software.
	The inputs consist of a mesh of finite elements representing the geometrical model, the governing partial differential equations describing the underlying laws of physics representing the physical model, and processing parameters (solver, time step, etc.).
	Although the geometrical model and the physical model appear to be equally important in their own rights \cite{zienkiewicz2005finite}, the process of defining a suitable mesh sometimes appears to dominate the overall process.
	During the processing step, the FEA simulation software transforms the geometrical model and the physical model into a system of algebraic equations, solves them according to the processing parameters, and reports the results back to the modeler, e.g., by visualizing the solution.
	Moreover, these steps are often executed repeatably until an acceptable solution, e.g., in terms of smoothness or convergence, is found.
	For example, incrementally refining the physical model and the mesh are considered to be crucial in determining whether the model has been built right (verification) and whether the right model has been built (validation)~\cite{hicks2015my}.
	
	In the field of FEA so far, workflows have been primarily focused on automating these steps (see Section \ref{sec:related-work}).
	However, they do not cover information about the simulation studies, e.g., the research question, assumptions, requirements, or further input data despite reporting guidelines suggest to make these accessible \cite{erdemir2012considerations,ABDELMEGID2020101175}.
	Moreover, they assume that the modeler can provide a geometrical model and physical model while ignoring the intricate processes involved in developing them and their relations to other products of the study, e.g., the input data used to develop the mesh or the requirements which were checked by the experiments.
	However, the information models of the artifacts within our artifact-based workflow already capture some meta-information about the simulation study while the lifecycle models are structuring and relating the development processes of the conceptual model, requirements, the simulation models, and experiments.
	Thus, adopting the artifact-based workflow for an entire FEA simulation study appears not only possible but also meaningful. 
	
	In contrast to stochastic discrete-event simulation studies, the FEA simulation model appears to consists of two parts, the geometrical model and the physical model, which were fused by specifying boundary conditions.
	Thus, the workflow has to be adapted to represent the geometrical model and the physical model as individual artifacts but also to treat them as constituents of the simulation model.
	This also implies that the milestones of \stage{Calibrating simulation model} and \stage{Validating simulation model} have to be adapted to be invalidated whenever the modeler revises the geometrical or physical model.
	
	Also, the data that is used as input gains weight in FEA studies, as these might be images that are transformed into the geometrical model and might make up a large part of developing the simulation model. Thus, input data have to be represented as an artifact to make their relations with the geometrical and physical model explicit.
	
	Finally, the FEA simulation studies rely on solvers that approximate the correct solution for the simulation. 
	Whereas in our workflow the simulation data produced by stochastic discrete-event simulation could be assumed to be correct (given a sound implementation of the simulation algorithm), in the case of approximative calculations a different approach is needed. 
	Now the simulation data need a life cycle of their own, i.e., stages that record whether the simulation data have been reproduced with a different simulation algorithm or a different mesh so to test the accuracy of the results. 
	In the latter case, the reproduction serves the verification of the geometrical model, i.e., whether the model has been built correctly.
	
	\subsection{Adaptation towards FEA Simulation Studies}
	\label{sec:workflow}
	To capture an entire FEA study, new artifacts, such as input data and simulation data are introduced, and the simulation model has to be redefined, as it now consists out of the geometrical model and physical model both forming artifacts of their own.
	
\begin{figure}
		\centering
		\includegraphics[width=.9\textwidth]{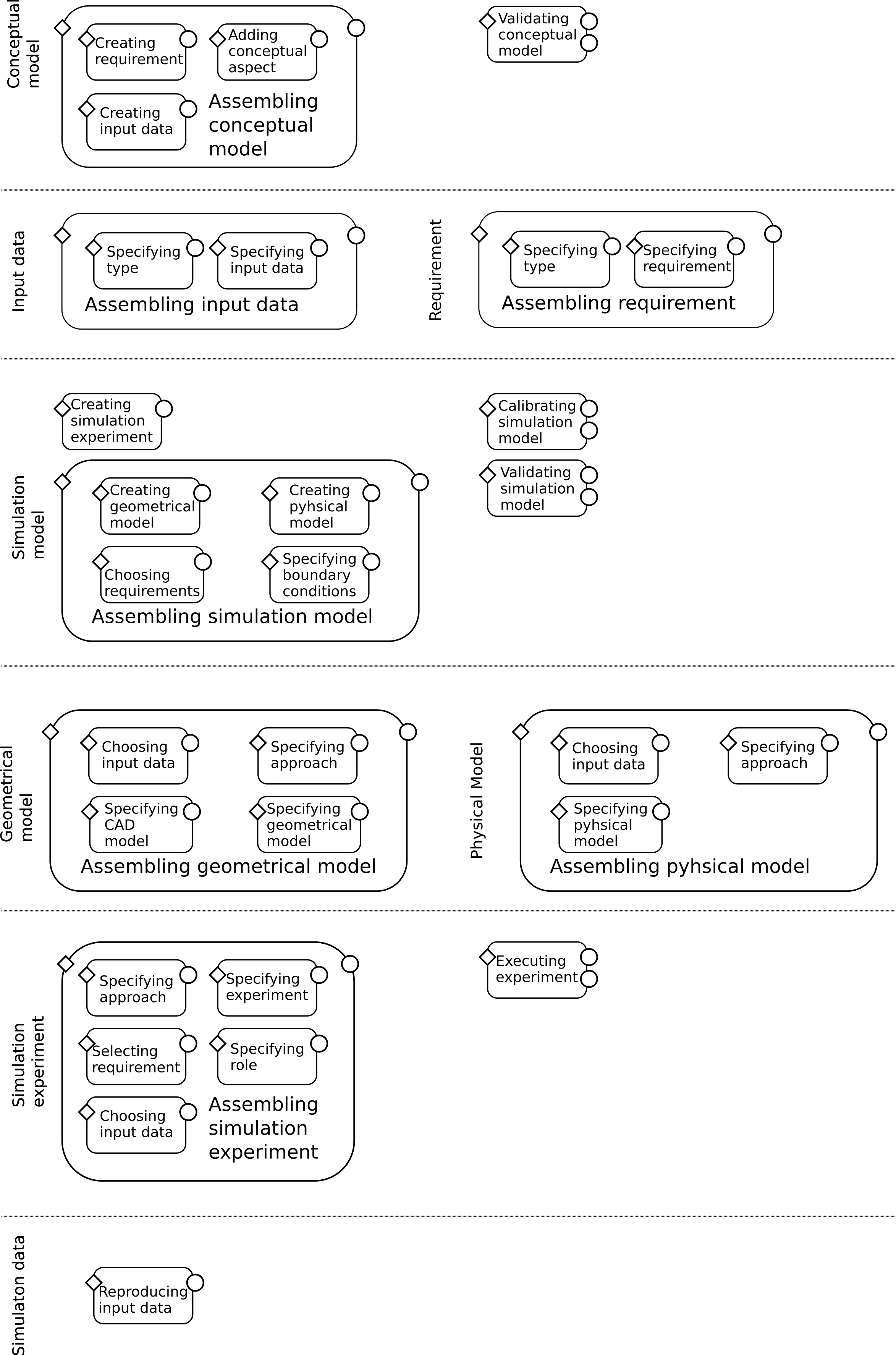}
		\caption{Overview of the adapted workflow. The new artifacts represent the input data, the physical model, the geometrical model, and the simulation data. The simulation model artifact has been revised, in particular, the assembling simulation model now consists of the stages to create a geometrical model, create a physical model, and specify boundary conditions. The Assembling stage of the conceptual model now includes also the creation of the input data artifact. The simulation data artifact is implicitly created by executing a simulation experiment.}
		\label{fig:adapt-artifact}
	\end{figure}

	Figure~\ref{fig:adapt-artifact} shows the artifact-based workflow for FEA simulation studies.
	The workflow consists of eight artifacts, four of which are new.
	Also, stages of the existing artifacts, such as the conceptual model, have to be adapted to integrate the new artifacts into the workflow, or to take the specifics of FEA workflows, e.g., the specification of boundary conditions, into account.
	
	To be clear, the artifact-based workflow for FEA simulation studies does not constrain the modeler to a specific tool, e.g., SALOME~\cite{SALOME} to generate the geometrical model or FEniCS~\cite{AlnaesBlechta2015a} to specify the physical model and solve the system of algebraic equations.
	Likewise, it does not support the user in applying these tools either.
	An implementation of the workflow might use these tools to implement \emph{toolboxes} to provide stage-specific editors allowing the modeler to provide the inputs for the information models, to check the syntactical correctness of specifications, and to execute the simulation experiments.
	Additional implementation details can be found in our original publication \cite{ruscheinski2019artifact}.
	
	\subsection{New Artifacts}
	The new artifacts for our workflow are input data, the physical model, the geometrical model, and the simulation data.
	
	\paragraph{Input Data Artifact}
	The input data is all the data used to assemble the physical model or the geometrical model.
	Further, input data might also be used in the simulation experiment to run simulations with different parameter values or over a range of parameter values.
	Consequently, the form of the input data might differ based on the application in the workflow, e.g., the input data for the geometrical model might comprise anatomical images from CT scans and/or might be a CAD model~\cite{KLUESS200923} based on which the mesh is created.
	The input data for the physical model and simulation experiments might refer, e.g., to measurements from laboratory experiments or other simulation studies~\cite{erdemir2019deciphering}.
	
	In our workflow, the input data artifact forms a new constituent of the conceptual model, as we use the conceptual model as a container for all information that helps us in developing the simulation model~\cite{Fujimoto2017}.
	Similar to the lifecycle of the requirement artifact, the lifecycle of the input data artifact consists of one main stage (\stage{Assembling input data}) in which the modeler has to specify the input data type before the input data are specified.
	The latter can also be a link to a data file containing the input.
	Further, we require that the modeler provides the source of the input data, e.g., by adding a DOI as part of the specification to ensure the provenance of the information (see Section~\ref{subsec:documentation}).
	
	After specifying the input data, the assembling stage can be left by the modeler.
	Thereby, the stage reaches its milestone which indicates that the assembling of the input data has been completed.
	
	\paragraph{Physical Model Artifact}
	The physical model describes the physical system, i.e., the physical laws, which should be used in the simulation to analyze the behavior of the geometrical model.
	The physical laws are typically specified using partial differential equations (PDEs) describing the time- and space-dependent behavior of the system.
	
	In the \stage{Assembling physical model}, the modeler can choose among the already assembled input data from the conceptual model to explicitly relate input data to the physical model.
	Further, the modeler has to specify a modeling approach, i.e., the physical modeling toolbox, before the physical model can be specified.
	After specifying the physical model, the assembling stage can be left and reaches its milestone to indicate that the physical model was successfully assembled.
	
	\paragraph{Geometrical Model Artifact} 
	The geometrical model specifies the spatial dimension of the simulation model.
	Therefore, the geometrical model forms a mesh, i.e., a subdivision of a continuous geometric space into discrete geometric and topological cells, containing information about shape and size, and spatially resolved information, e.g., referring to material.
	As part of the simulation model, the geometrical model serves as the initial state, based on which the dynamics evolve due to the constraints given by the physical model.
	
	The development process of the geometrical model is encapsulated by the \stage{Assembling geometrical model}.
	Within this stage, the modeler has to specify a modeling approach, i.e., the geometrical modeling toolbox, providing the means to create a mesh from a CAD model.
	Before the mesh can be created, the modeler has to choose the input data used to create the mesh, e.g., the images from CT scans~\cite{KLUESS200923}, and specify a CAD model, e.g., by importing an existing model or creating a new one.
	After completing the meshing process, the assembling stage can be left by the modeler.
	
	\paragraph{Simulation Data Artifact}
	The simulation data artifact represents the data generated by executing simulation experiments.
	Depending on the experiment, the data might consist of validation results, optimization results, or data which is further analyzed, e.g., by visualization.
	Reproducing the simulation data by using different solvers, etc. increases the trust in the data and also ensures their reproducibility.
	
	The artifact has only one stage, i.e., \stage{Reproducing simulation data} with two possible outcomes, fail and succeed.
	By entering the stage, we automatically generate a new experiment by adapting the original simulation experiment to check whether the simulation data can be reproduced.
	Therefore, we allow the modeler to manually update the generated experiment to specify, e.g., another solver or a newly meshed geometrical model.
	After that, the experiment can be executed to generate new data which is compared, based on some metrics, with the previously generated data.
	Depending on the result, one of the two milestones is reached to indicate that the data could or could not be reproduced.
	
	\subsection{Integrating the new artifacts}
	To integrate the new artifacts into the workflow, we revise the existing artifact lifecycles to enable the creation of the new artifacts and interaction between artifacts during the workflow.
	Therefore, we will add new stages and revise the guards and milestones of existing stages in our workflow.
	
	\paragraph{Integrating the Input Data Artifact}
	As also suggested in~\cite{Robinson2008b}
	the input data shall be part of the conceptual model.
	Therefore now, the \stage{Assembling conceptual model} stage comprises a stage for creating a new input data artifact.
	Consequently, all created input data artifacts are stored as part of the information model of the conceptual model artifact.
	To ensure the coherence of the information in the conceptual model, we further revise the guard and the milestones of the \stage{Validating conceptual model}.
	
	First, we only allow to enter the validating stage when all created input data artifacts are assembled, i.e., fully specified.
	Therefore, the guard is extended to check whether
	all created input data artifacts have reached the milestone of their assembling stage.
	Further, we revise the milestones of the validating stage to be invalidated whenever a created input data artifact is modified.
	
	Since often simulation experiments make use of inputs, we also extend the \stage{Assembling simulation experiment} to enable the modeler to choose from the input data artifacts.
	The references for the chosen input data artifacts are also stored in the information model of the simulation experiment artifact.
	
	\paragraph{Integrating the Physical Model and the Geometrical Model}
	The simulation model of FEA simulation studies consists of two parts: the physical model and the geometrical model.
	To represent this in our artifact-based workflow, we need a different simulation model artifact.
	Therefore, we removed the \stage{Specifying modeling formalism} and the \stage{Specifying simulation model} from the \stage{Assembling simulation model} since the modeling formalism and the specification of the model are now handled in the \stage{Assembling physical model}.
	Further, we allow creating a single physical model and a single geometrical model during the assembling of the simulation model.
	Consequently, the references to the created artifacts are stored as part of the information model of the simulation model artifact.
	As before, each simulation study might contain multiple simulation model instances, each with its own information model.
	
	The \stage{Specifying boundary condition} stage allows the user to specify the boundary conditions for the physical model given a geometrical model.
	Consequently, the geometrical model and the physical model has to be assembled before the boundary conditions can be specified.
	Moreover, the modeler can only leave the \stage{Assembling simulation model} after the boundary conditions have been specified to ensure that the simulation model can be used in the simulation experiments.
	
	The milestones of the \stage{Validating simulation model} are invalidated whenever the geometrical or physical model of the simulation is modified or the boundary conditions are changed to ensure that the changed simulation model will be validated again.
	Note that also experiments conducted in the \stage{Calibrating simulation model} may refer to either the geometrical model (e.g., finding a mesh where the solution converges) or the physical model (e.g., fitting the material properties to achieve a certain electric field distribution).
	
	\paragraph{Integrating the Simulation Data}
	Unlike previous artifacts, the simulation data artifacts are not created explicitly by a modeler via the \stage{Creating-x} of some artifact but are created implicitly by entering the \stage{Executing experiment}, the \stage{Validating simulation model} or the \stage{Calibrating simulation model stage} of the simulation experiment artifact.
	
	\section{Supporting of FEA Simulation Studies}
	\label{sec:case-study}
	In the following, we will demonstrate how a modeler moves through the artifact-based workflow based on an exemplary FEA simulation study.
	Also, we will show how the artifact-based workflow can be exploited to guide the modeler through the simulation study while documenting the development processes of the different work products, and how we can automatically generate simulation experiments.
	
	\subsection{Case Study: Simulating an Electrical Stimulation Chamber}
	In electrical stimulation (ES), it is very important to characterize the local effect of the applied stimulation.
	For that, we carry out field simulations to go beyond simplistic analytical estimates.
	As FEA is a method that is capable of treating complex geometries, it is the method of choice.
	
	The goal of the case study is to compute the electric field (EF) distribution in an ES chamber \cite{Mobini2016, Mobini2017} following the workflow published in~\cite{Budde2019}.
	Therefore, the modeler begins with collecting information about the geometry of the chamber, the spacing and material of the electrodes, and the cell culture medium.
	Based on this, the modeler continues by developing the geometrical model of the chamber using a CAD model to generate a mesh.
	Finally, the modeler specifies a physical model describing electrical field distribution based on Maxwell's equations, specifies the boundary conditions on the mesh, and validates the simulation model.
	
	\paragraph{Step 1: Collecting Information}
	The modeler begins with collecting the information about the spacing of the two electrodes ($\SI{22}{\mm}$) as well as their general shape (bent, L-shaped platinum wires) from the publications \cite{Mobini2016, Mobini2017}.
	Moreover, the modeler finds that standard 6-well dishes are used and assumes that the volume of the cell culture medium is approximately the recommended volume of $\SI{3}{\ml}$~\cite{TechnoPlasticProducts2018Basics}.
	
\begin{figure}[!ht]
		\begin{subfigure}{0.5\textwidth}
			\centering
			\includegraphics[height=4cm]{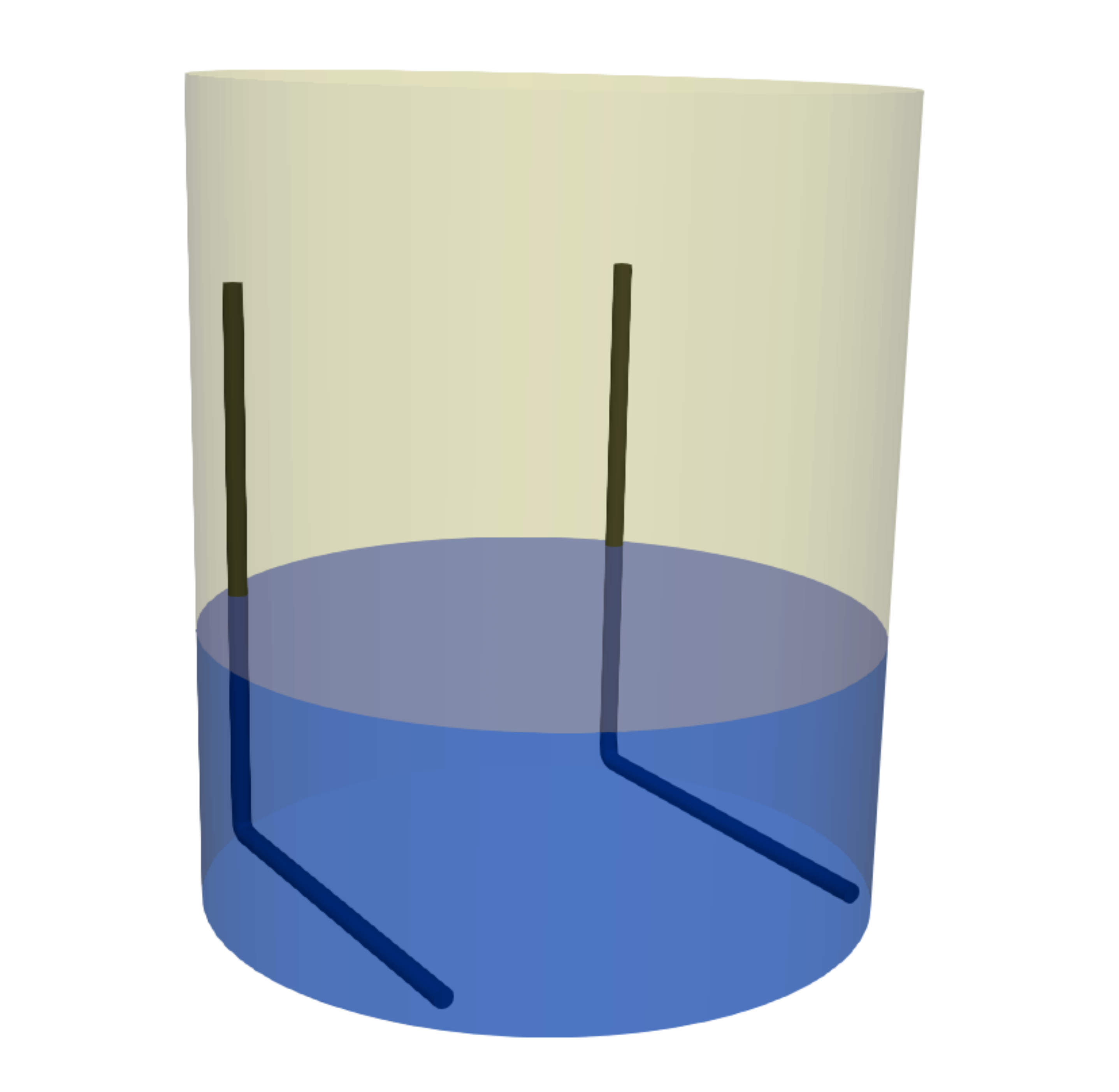}
			\caption{CAD Model}
			\label{fig:cad-model}
		\end{subfigure}
		\begin{subfigure}{0.5\textwidth}
			\centering
			\includegraphics[height=4cm]{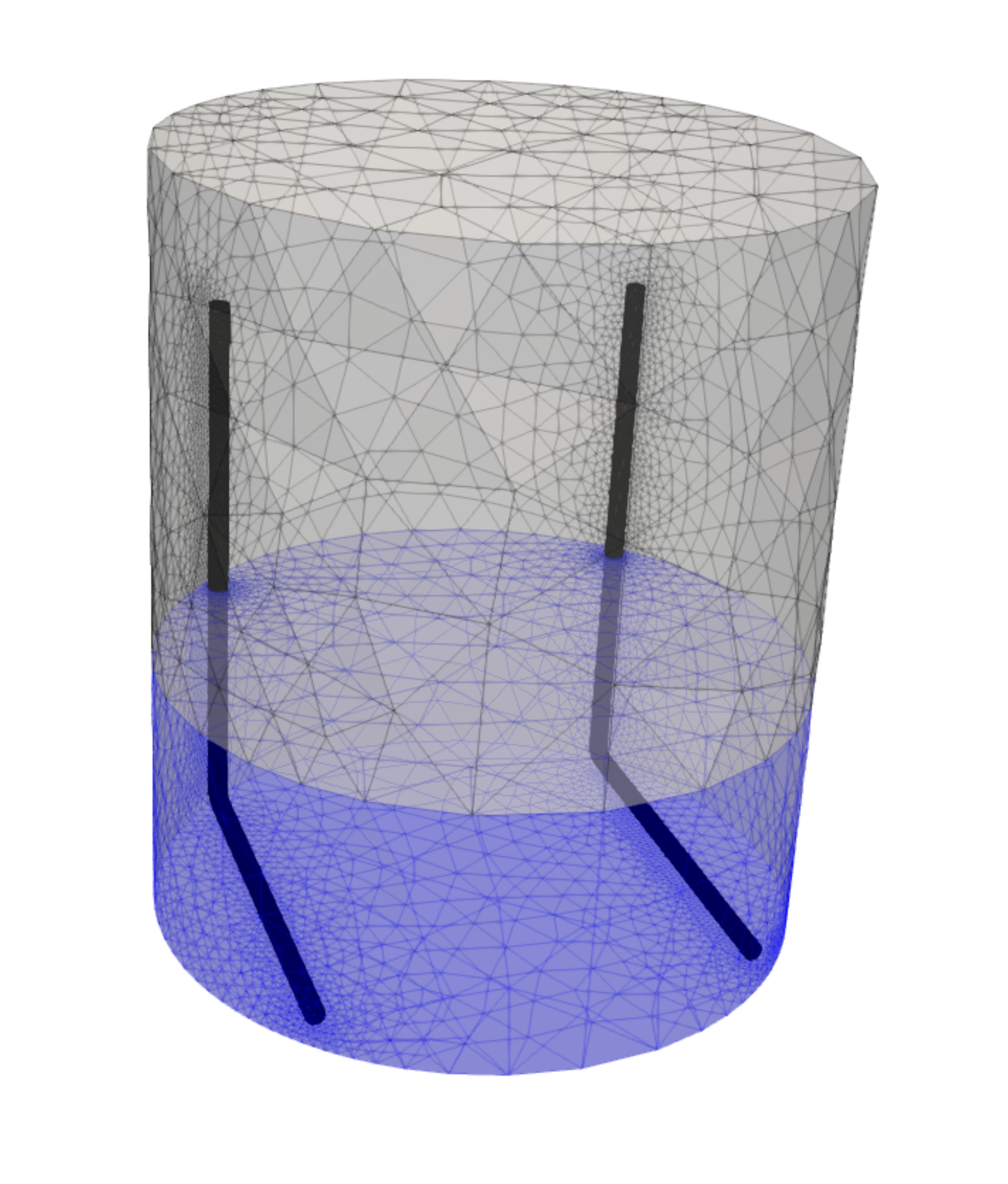}
			\caption{Mesh}
			\label{fig:cad-mesh}
		\end{subfigure}
		\caption{CAD-Model and mesh of ES chamber with the cell culture medium (blue), the two electrodes (black) and air (shaded).}
	\end{figure}

	\paragraph{Step 2: Developing the Geometrical Model}
	For the geometrical model of the ES chamber, the modeler creates a new CAD model in SALOME~\cite{SALOME}.
	Using the information about the inner diameter of the well, the modeler computes the fill level and creates a geometry object representing the cell culture medium in the CAD model.
	To fit the electrodes into the well at the given spacing of $\SI{22}{\mm}$, assumptions are made on the length of the vertical part ($\SI{22}{\mm}$) and length of the horizontal part ($\SI{28}{\mm}$) of the electrode.
	In the next step, the cylindrical electrodes are created in the CAD model and placed as described previously.
	The space between the medium and the highest point of the electrodes is filled with air.
	Finally, the modeler cuts the electrodes out of the geometry as only the EF in the cell culture medium is of interest.
	The final CAD model is shown in Figure~\ref{fig:cad-model}.
	
	Using the CAD model, a mesh consisting of tetrahedral elements (see Figure~\ref{fig:cad-mesh}) is generated in SALOME using the Netgen mesh generator~\cite{Schoberl1997}.
	Using \enquote{a priori} knowledge that the mesh needs to have a higher resolution on the electrode surfaces to resolve to geometry well, the choice of meshing parameters is influenced.
	Afterward, the modeler adds the material information about the electrical conductivity of the air and cell culture medium to the respective elements of the mesh.
	Likewise, the electrode surfaces are marked where the electrical potentials will be enforced by the specification of the boundary values.
	The quality of the mesh is checked visually and by the meshing algorithm.
	Finally, the resulting mesh with the material information and the marked electrode surfaces is exported using the MED-format supported by SALOME.
	
	\paragraph{Step 3: Developing the Physical Model}
	To compute the EF distribution in the ES chamber, the modeler has to develop a suitable physical model.
	As the ES is carried out by a direct current, the electrostatic representation of Maxwell's equations is an appropriate modeling choice (see Equation~\ref{eq:maxwell})~\cite{Malmivuo1995}.
	In this case, the electric field (see Equation~\ref{eq:field}) can be computed using the following equations
	
	\begin{subequations}
		\begin{align}
		\Delta \Phi = 0 \enspace, \label{eq:maxwell} \\
		\mathbf{E} = - \nabla \Phi \label{eq:field}
		\end{align}
	\end{subequations}
	where $\Phi$ is the electric potential.
	
	To solve the direct current problem for different materials, the modeler employs the low-frequency limit of the electrostatic equation to include the materials' electrical conductivities $\sigma(\mathbf{r})$~\cite{Malmivuo1995,Budde2019}.
	
	Thus, we use
	\begin{equation}
	\label{eq:EQS}
	\nabla \cdot \left( \sigma \nabla \Phi \right) = 0
	\end{equation}
	instead of Equation~\ref{eq:maxwell}.
	
	To solve these equations, the boundary values for the voltage applied at the two electrodes, i.e., the electric potential has to be specified.
	In the original study~\cite{Mobini2017}, no voltage is reported but the preceding study~\cite{Mobini2016} mentions $\SI{2.2}{\V}$.
	However, due to the linearity of Equation~\ref{eq:EQS} the boundary values can be set arbitrarily since solutions can be scaled in a post-processing step to match reality.
	Finally, the information about the conductivities can be determined experimentally or extracted from a database such as ITIS\cite{itisDB}.
	
	Using this information, the modeler specifies the physical model using the EMStimTools Python package (see Code Listing~\ref{lst:physo-mo}) providing a solution template for the Equations~\ref{eq:field} and~\ref{eq:EQS}.
	
	\begin{figure} [h]
		\begin{lstlisting}[caption = Specification of the pyhsical model in EMStimTools., label=lst:physo-mo, numbers=left, autogobble=true, frame=single,language=yaml,columns=flexible,aboveskip=3mm, belowskip=3mm]
		# Using the electrical stimulation template
		physics: ES
		
		# Specifying the materials
		materials: [Air, Medium]
		
		# Specifying the material conductivities in S/m
		conductivity:
		Air: 1e-14
		Medium : 1.0
		
		# Specifying the boundary conditions and boundary values
		boundaries:
		Dirichlet:
		Contact1: 1.0
		Contact2: 0.0
		\end{lstlisting}
	\end{figure}
	
	\paragraph{Step 4: Validating the approach and the results}
	To run a simulation experiment, the modeler uses the Python programming language to interface EMStimTools.
	Therefore, EMStimTools loads the meshed geometry and the physical model based on which a linear system of equations can be generated and be solved for the electric potential by FEniCS \cite{AlnaesBlechta2015a, Logg2012} routines.
	Afterward, the results are passed back to the Python environment for further post-processing, i.e., visualizing the results via ParaView~\cite{ParaView}.
	
\begin{figure}[!h]
		\centering
		\includegraphics[width=.8\textwidth]{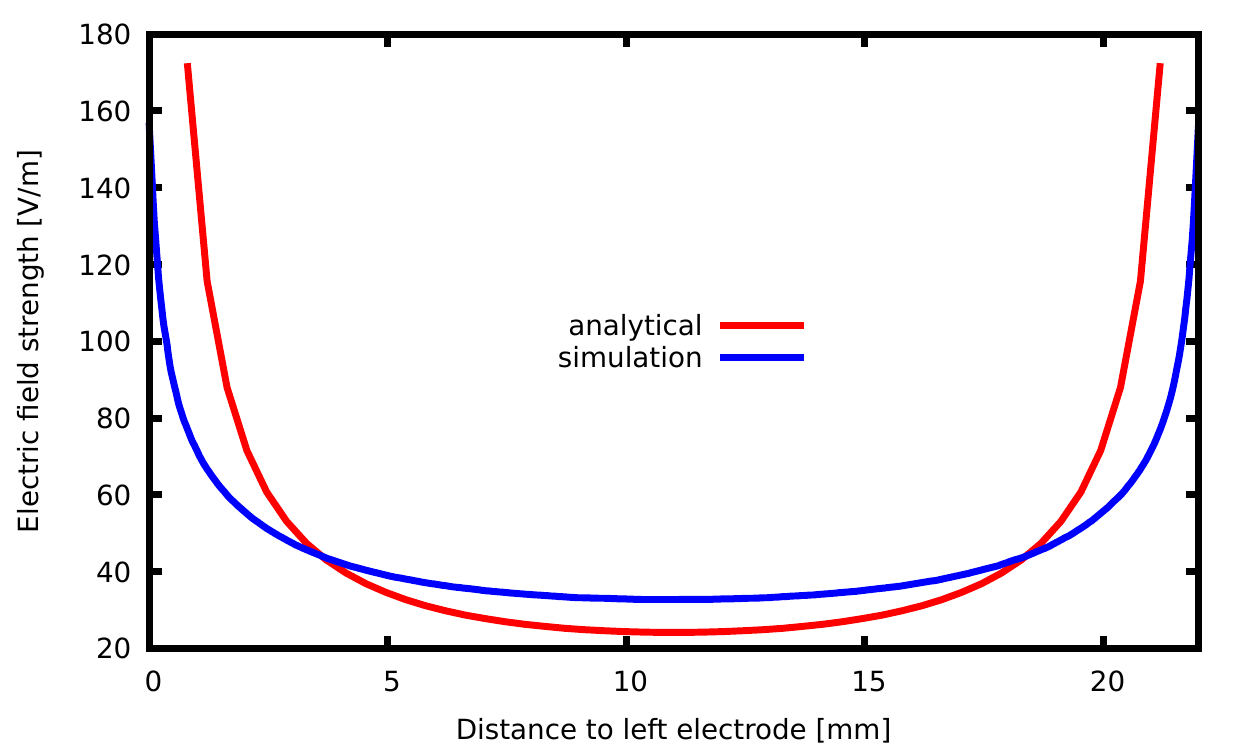}
		\caption{Electric field strength between the electrodes.
			The left electrode is located at $y=\SI{-11}{\mm}$.
			The analytical result is based on~\cite{Hronik-Tupaj2011}.
		}
		\label{fig:field_center_mobini}
	\end{figure}

	The validity of the approach and the results can be substantiated by comparing them to analytical functions or by applying methods such as adaptive refinement based on error estimators~\cite{Rognes2013}.
	A thorough convergence study using global mesh refinement, i.e., splitting each tetrahedron in two or four smaller ones, could be attempted but might lead to a huge amount of degrees of freedom.
	Hence, the modeler compares the result first with an analytical solution of a similar problem~\cite{Hronik-Tupaj2011, Budde2019} (see Fig.\ref{fig:field_center_mobini}).
	Note that the analytical solution does not correspond to the mathematically exact solution of the problem. 
	Such a solution cannot be obtained analytically for the given geometry.
	
\begin{figure}[!h]
		\centering
		\includegraphics[width=.8\textwidth]{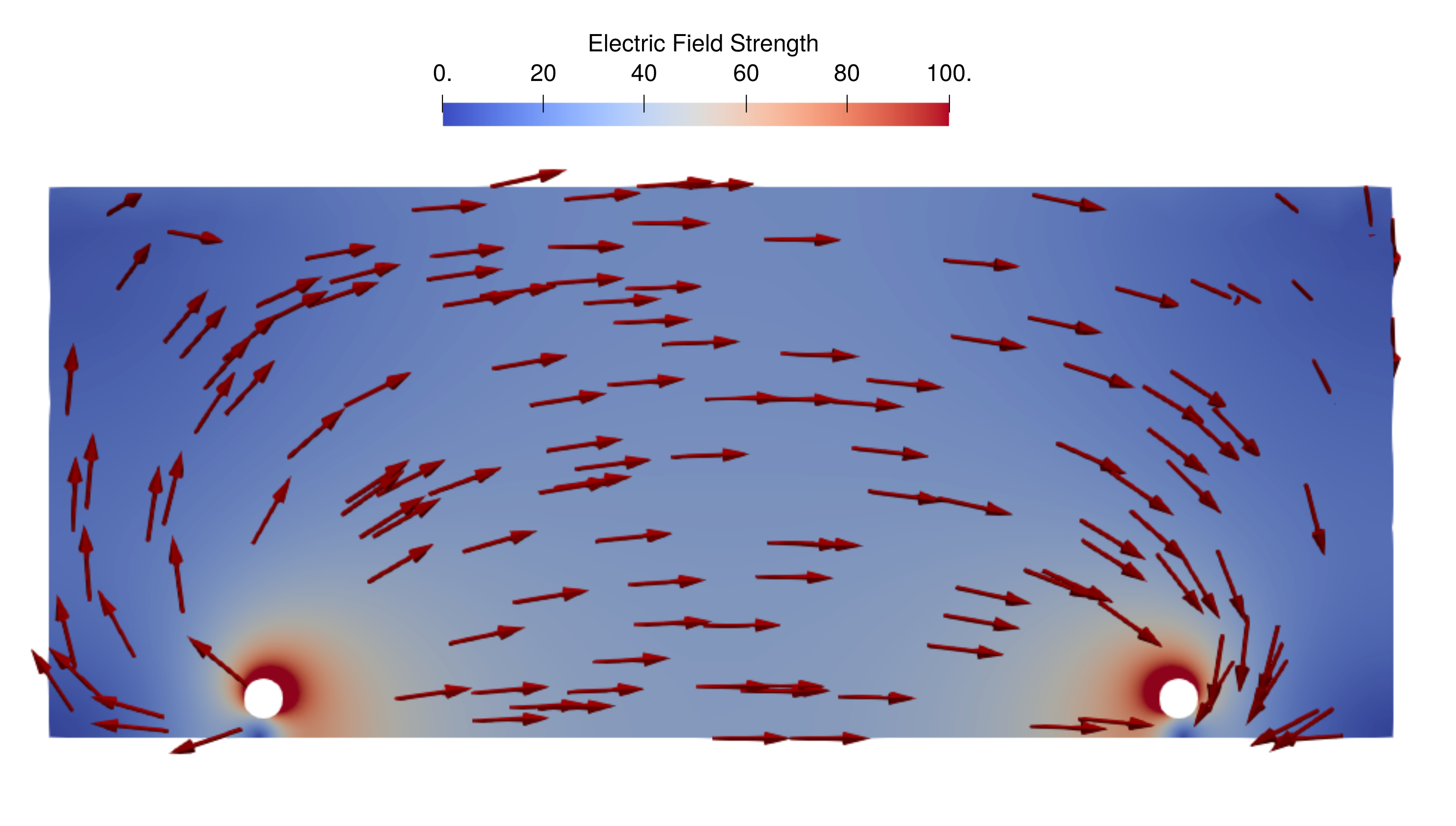}
		\caption{Electric field distribution in the center of the well using the ES chamber described in~\cite{Mobini2017}. The field strength is shown together with the orientation to highlight the field homogeneity between the two electrodes.}
		\label{fig:field_wires_mobini}
	\end{figure}

	Afterward, the modeler checks the convergence by observing the solution in the critical area around the electrode and manually refines the mesh step-by-step.
	Hereby, the modeler pays attention to resolving the curved shape and the field enhancement at the edges.
	Finally, after the modeler established that the model is valid, the EF distribution of the EM chamber is computed and visualized (see Figure~\ref{fig:field_wires_mobini}).

	\subsection{Running the Case Study in the Artifact-based Workflow}
	Here, we show how the four steps of the case study are represented by the workflow.
	Our workflow starts with an empty conceptual model artifact. 
	
\begin{figure}[h]
		\includegraphics[height=3cm]{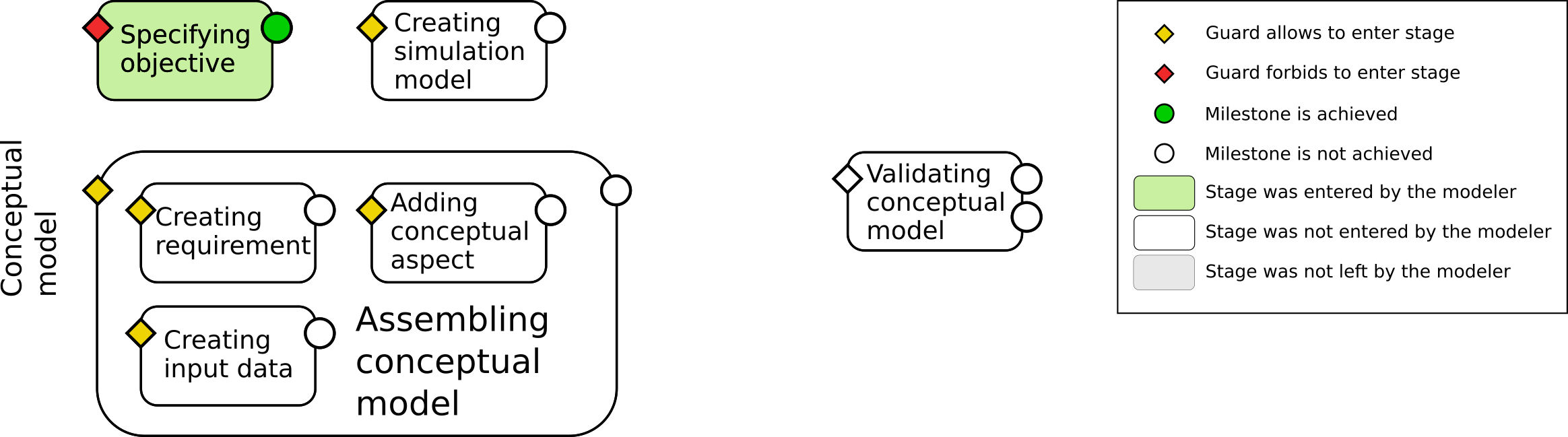}
		\caption{Overview of the lifecycle of the conceptual model artifact after specifying the objective, all but the \stage{Specifying objective} are enabled.}
		\label{fig:case-study-01}
	\end{figure}

	\paragraph{Step 1: Assembling the Conceptual Model}
	First, the modeler enters the \stage{Specifying objective} to provide information about the objective of the planned simulation study.
	After finishing the specification, the objective is stored in the information model of the artifact (see Table~\ref{table:cmo-obj}) and the stage achieves its milestone.
	This triggers the re-evaluation of the guards of the stages.
	The evaluation of the \stage{Specifying objective} guard is set to false since the objective was already specified but the guards of the other stages are evaluated to true and therefore allow the modeler to continue with any of the respective stages (see Figure~\ref{fig:case-study-01}).
	
	\begin{table}[h]
		\caption{Overview of the information model of the conceptual model artifact after specifying the objective.}
		\label{table:cmo-obj}
		\begin{tabularx}{\textwidth}{llX}
			\toprule
			Name & Type & Description \\
			\midrule
			objective & Text & Compute the electric field distribution in an electrical stimulation chamber previously described in~\cite{Mobini2017}\\
			\bottomrule
		\end{tabularx}
	\end{table}

	The modeler continues by specifying the information about the ES chamber.
	Therefore, the \stage{Creating input data} is entered multiple times to create new input data artifact instances for the details about the chamber.
	Subsequently, the input data artifact instances are assembled by specifying the type and the information about the chamber
	(see Table~\ref{table:input-data}).
	
	\begin{table}[h]
		\caption{Overview of the information models of the assembled input data artifacts.}
		\label{table:input-data}
		\begin{tabularx}{\textwidth}{Xlll}
			\toprule 
			Name                          & Type & Specification  & Source                                             \\ 
			\midrule
			Electrode - Shape             & Text & bent, L-shaped & Publications~\cite{Mobini2016,Mobini2017}          \\
			Electrode - Material          & Text & platinum       & Publications~\cite{Mobini2016,Mobini2017}          \\
			Electrode - Vertical length   & Data & $\SI{22}{\mm}$ & Publications~\cite{Mobini2016,Mobini2017}          \\
			Electrode - Horizontal length & Data & $\SI{28}{\mm}$ & Publications~\cite{Mobini2016,Mobini2017}           \\
			Dish - Material               & Text & plastic        & Publications~\cite{TechnoPlasticProducts2018Basics} \\
			Dish - Size                   & Text & 6-well         & Publications~\cite{TechnoPlasticProducts2018Basics} \\
			Medium - Volume               & Data & $\SI{3}{\ml}$  & Publications~\cite{TechnoPlasticProducts2018Basics} \\
			Medium - Electrical conductivity         & Data & $\SI{1}{\siemens \per \meter}$              & Laboratory measurement \\
			Air - Electrical conductivity & Data & $\SI{1E-14}{\siemens \per \meter}$ &  ITIS~\cite{itisDB} \\
			\bottomrule
		\end{tabularx}
	\end{table}
	
	After specifying the type in the \stage{Specifying type}, the corresponding guard prevents the modeler from re-entering the stage.
	The modeler finishes the collection of information by leaving the \stage{Assembling conceptual model} of the conceptual model artifact.
	By doing that the corresponding milestone of the stage is achieved.
	
\begin{figure}[!h]
		\includegraphics[height=6cm,width=\textwidth]{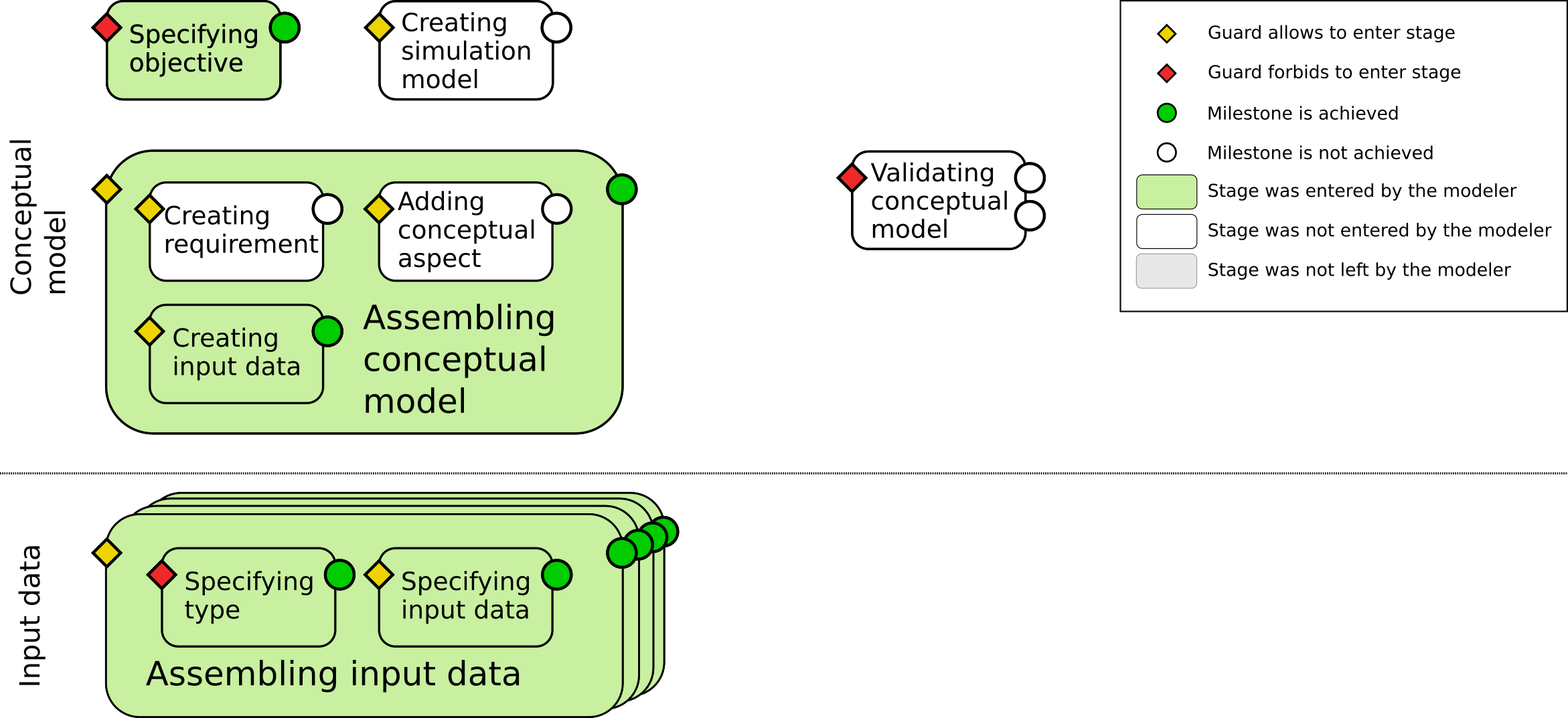}
		\caption{Overview of the artifact livecycles after creating and assembling the input data artifacts and leaving the \stage{Assembling conceptual model}.}
		\label{fig:case-study-02}
	\end{figure}

	At this point,  the guards allow the modeler to continue by creating a simulation model artifact, re-entering the \stage{Assembling conceptual model} to collect more information, or validating the conceptual model (see Figure~\ref{fig:case-study-02}).
	
	\paragraph{Step 2: Assembling the Geometrical Model}
	The modeler continues in the workflow by creating a new simulation model artifact.
	In the \stage{Assembling simulation model}, the modeler enters the \stage{Creating geometrical model} to create a new geometrical model artifact representing the ES chamber.
	To reach the corresponding milestone of the \stage{Assembling simulation model} requires that a geometrical model and a physical model have to be created and assembled and that the boundary conditions have to be specified.
	The modeler continues to assemble the geometrical model.
	
	After entering the \stage{Assembling geometrical model}, the modeler can specify the geometrical modeling approach, i.e., geometrical modeling toolbox, or can choose the input data for the geometrical model.
	To achieve the milestone of this stage, first the modeler needs to define the geometrical model approach and a CAD model.
	After identifying SALOME as the geometrical modeling approach the milestone of the \stage{Specifying approach} is achieved.
	Thus, the evaluation of the guards of the geometrical model artifact now prevents the modeler from re-entering the \stage{Specifying approach} but still allows choosing the input data and also to define the CAD model.
	The modeler chooses the input data, i.e., the information about the electrodes, the well, and the medium, for the geometrical model.
	The references to the input data are stored in the information model of the geometrical model artifact to make their relation explicit.
	After completing the CAD model, the resulting CAD-file is passed from the geometrical modeling toolbox to the workflow environment to be stored in the information model of the geometrical model artifact (see Table~\ref{table:gmo-data}).
	
	\begin{table}[h]
		\caption{Information model of the geometrical model artifact.}
		\label{table:gmo-data}
		\begin{tabularx}{\textwidth}{lp{2cm}X}
			\toprule Attribute     & Type                         & Value                                                                                                                                                  \\\midrule
			inputs        & Input Data                   & \{Electrode - Shape, Electrode - Material, Electrode - Vertical length, Electrode - Horizontal length, Dish - Material, Dish - Size, Medium - Volume\} \\
			approach      & Geometrical-Modeling Toolbox & SALOME                                                                                                                                                 \\
			cad           & CAD-File                     & \texttt{Mesh\_chamber.step}                                                                                                                                    \\
			specification & Mesh-File                    & \texttt{Mesh\_chamber.med} \\
			\bottomrule
		\end{tabularx}
	\end{table}
	
\begin{figure}
		\includegraphics[width=\textwidth]{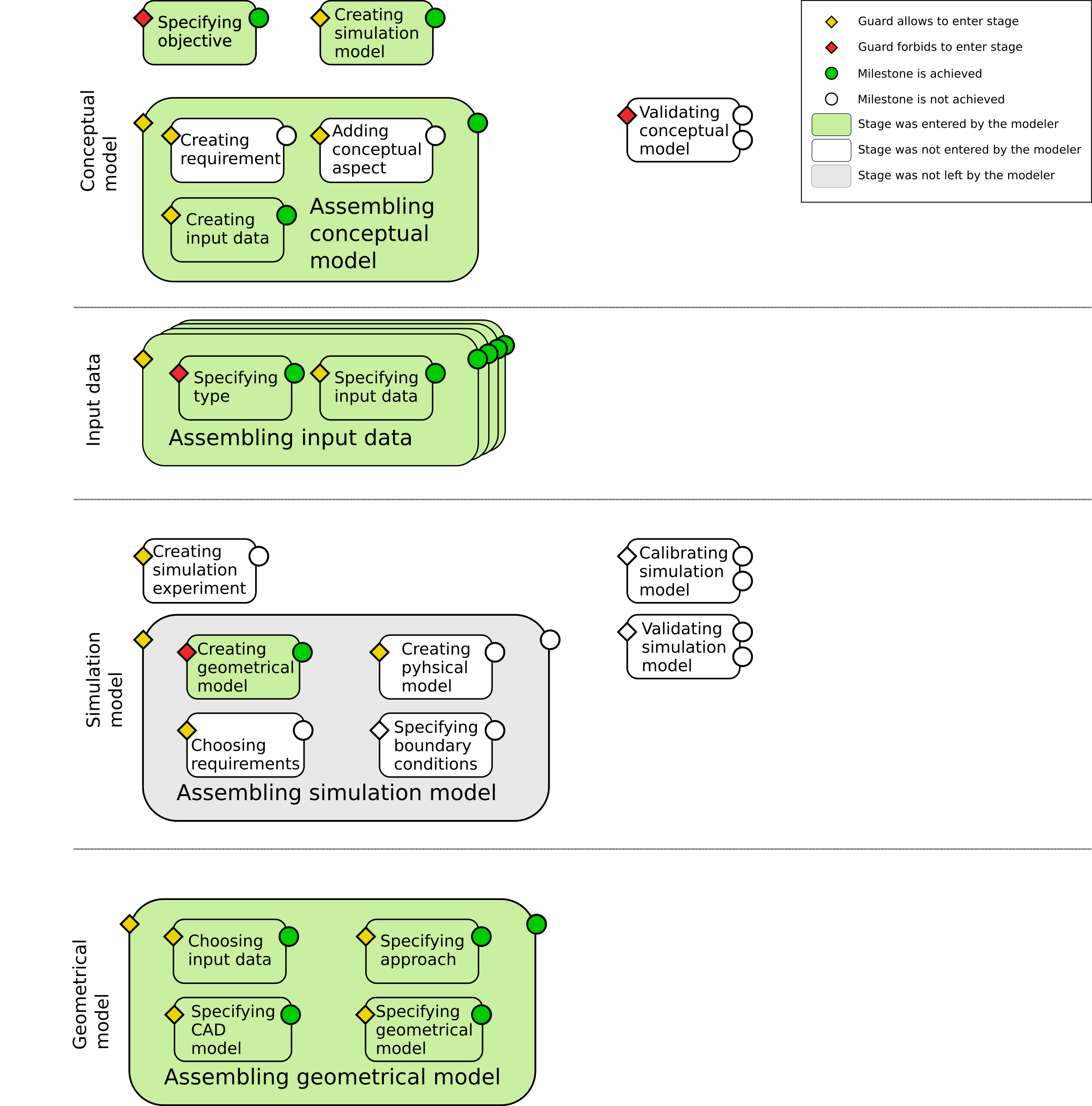}
		\caption{Overview of artifact' lifecycles after assembling the geometrical model.}
		\label{fig:case-study3}
	\end{figure}

	Now the re-evaluation of the guards also allows creating a mesh from the CAD model in the \stage{Specifying geometrical model}.
	The created mesh is also passed to the workflow environment to be stored in the information model.
	Finally, the modeler leaves the assembling stage by the corresponding milestone being achieved (see Figure~\ref{fig:case-study3}).

	\paragraph{Step 3: Assembling the Physical Model}
	After creating and assembling the geometrical model, the modeler continues in the \stage{Assembling simulation model} by creating a physical model artifact.
	
	In the \stage{Assembling physical model}, the modeler can choose the input data for the physical model or specify the physical modeling approach.
	Similar to the assembling stage of the geometrical model, the milestone of the \stage{Assembling physical model} stage requires information about the modeling approach in which the physical model will be specified.
	The modeler chooses the input data representing the material conductivities as input data to make the relation between those explicit. 
	In this case EMStimTools is entered as a modeling approach. 
	Lines 1--10 of the Code Listing~\ref{lst:physo-mo} form the physical model and are, correspondingly, stored in the information model of the artifact physical model.  
	
	\begin{table}[h]
		\caption{Information model of the physical model artifact.}
		\label{table:phymo-data}
		\begin{tabularx}{\textwidth}{llX}
			\toprule Attribute     & Type                      & Value                                   \\\midrule
			approach      & Physical-Modeling Toolbox & EMStimTools                             \\
			inputs        & \{Input Data\}            & \{Medium - Electrical conductivity, Air - Electrical conductivity\}                                    \\
			specification & Text                      & (see Code Listing~\ref{lst:physo-mo} lines 1 -- 10) \\
			\bottomrule
		\end{tabularx}
	\end{table}
	
\begin{figure}[!h]
		\includegraphics[width=\textwidth]{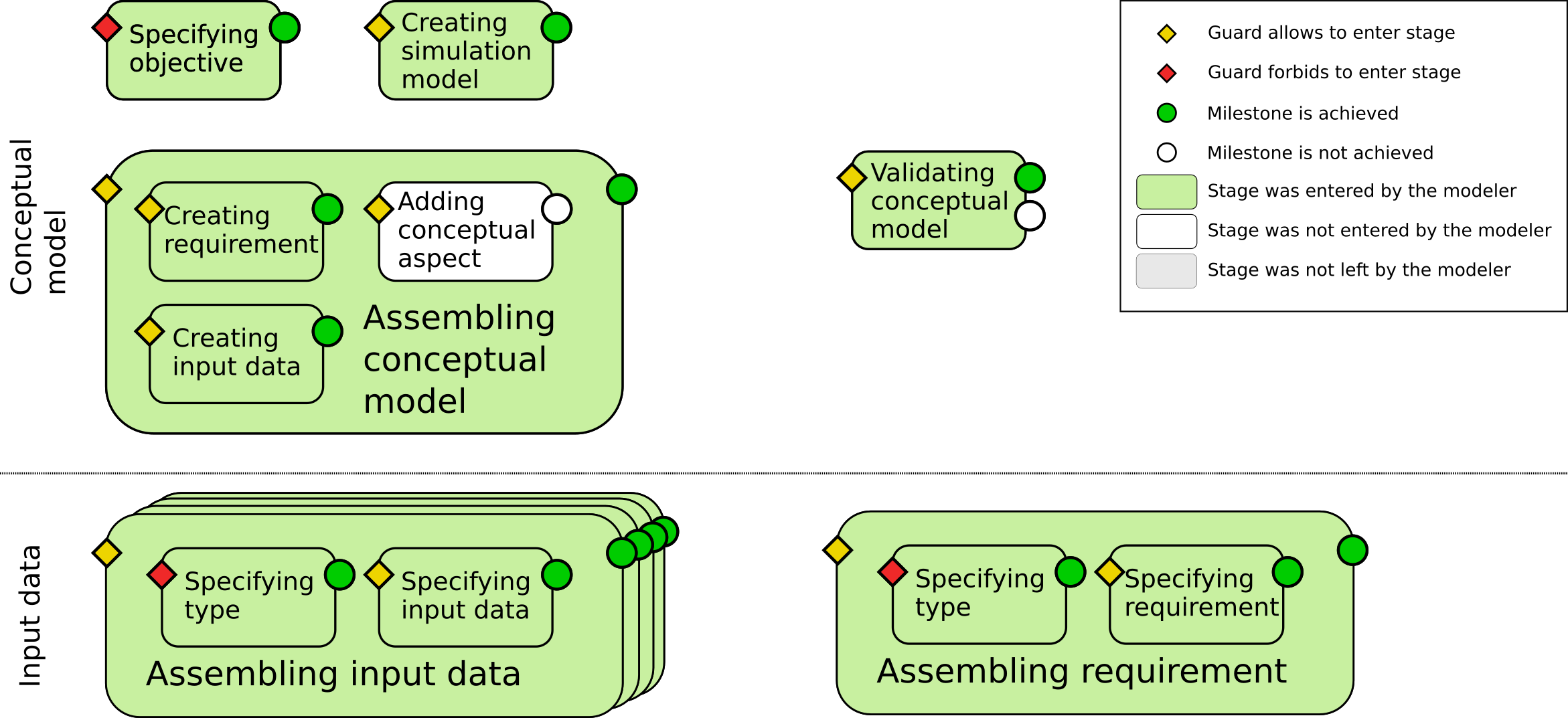}
		\caption{Overview of the life cycles of the conceptual model, requirement and input data.}
		\label{fig:case-study-04}
	\end{figure}

	Back in the \stage{Assembling simulation model}, the evaluation of the guards now allows us to specify the boundary conditions (see Figure~\ref{fig:case-study-04}).
	
	\begin{figure}[h]
		\includegraphics[width=\textwidth]{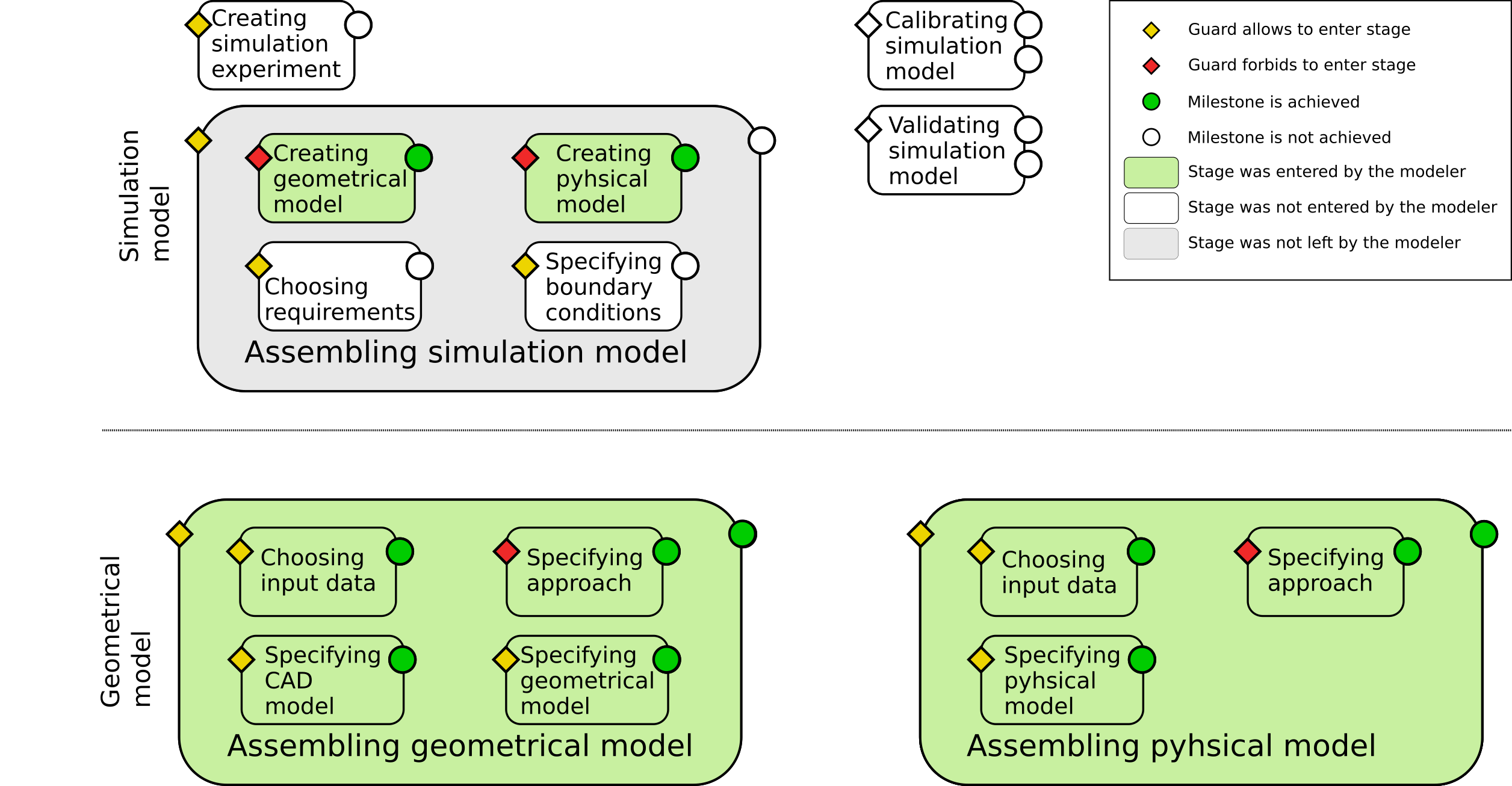}
		\caption{Overview of the lifecycles of the simulation model, geometrical model, and physical model.}
		\ref{fig:case-study-04}
	\end{figure}

	In the \stage{Specifying boundary conditions}, the YAML-editor of EMStimTools is used to define
	the boundary conditions (see Code Listing~\ref{lst:physo-mo} lines 13--16).
	The resulting boundary conditions are stored in the information model of the simulation model artifact.
	
	Finally, the modeler leaves the assembling-stage by which the milestone is achieved.
	In the workflow, the milestone serves as an indicator that the simulation model, i.e., the combination of the physical model and geometrical model, is ready to be used in simulation experiments.
	
	\paragraph{Step 4: Validating the Simulation Model}
	The modeler continues in the workflow to validate the simulation model.
	However, to enter the \stage{Validating simulation model} the guard requires that the modeler has  \begin{inparaenum} \item to create and assemble a requirement artifact specifying the property that should be checked, i.e., the analytical solution of the similar problem, \item to assign the requirement to the simulation model, \item to validate the conceptual model to ensure that the collected information about the ES chamber is consistent, and \item to assemble a validation experiment.
	\end{inparaenum}
	
	To create an artifact requirement, the modeler revisits the conceptual model artifact and enters the \stage{Creating requirement}.
	After this, the modeler enters the \stage{Assembling requirement} to assemble a requirement artifact representing the analytical solution of the similar problem.
	Before the requirement can be specified, the modeler has to specify the requirement type, i.e., data or logic formula.
	After specifying \enquote{data} as the requirement type and specifying the analytical solution, the modeler leaves the assembling stage.
	
	Back in the conceptual model artifact the evaluation of the guards now allows entering the \stage{Validating conceptual model}.
	After face validating, i.e., manually reviewing, all the collected information the modeler marks the conceptual model as valid.
	Based on this decision, the stage reaches the \emph{valid}-milestone (see Figure~\ref{fig:case-study-04}).
	
	After these steps, the modeler continues in the simulation model artifact by entering the \stage{Choosing requirements}.
	In the stage, the modeler selects the analytical solution which is afterward stored as a reference in the simulation model artifact.
	Next, a new simulation experiment artifact with role \enquote{validation} is created by entering the \stage{Creating simulation experiment} of the simulation model artifact.
	
\begin{figure}[!h]
		\centering
		\includegraphics[scale=0.4]{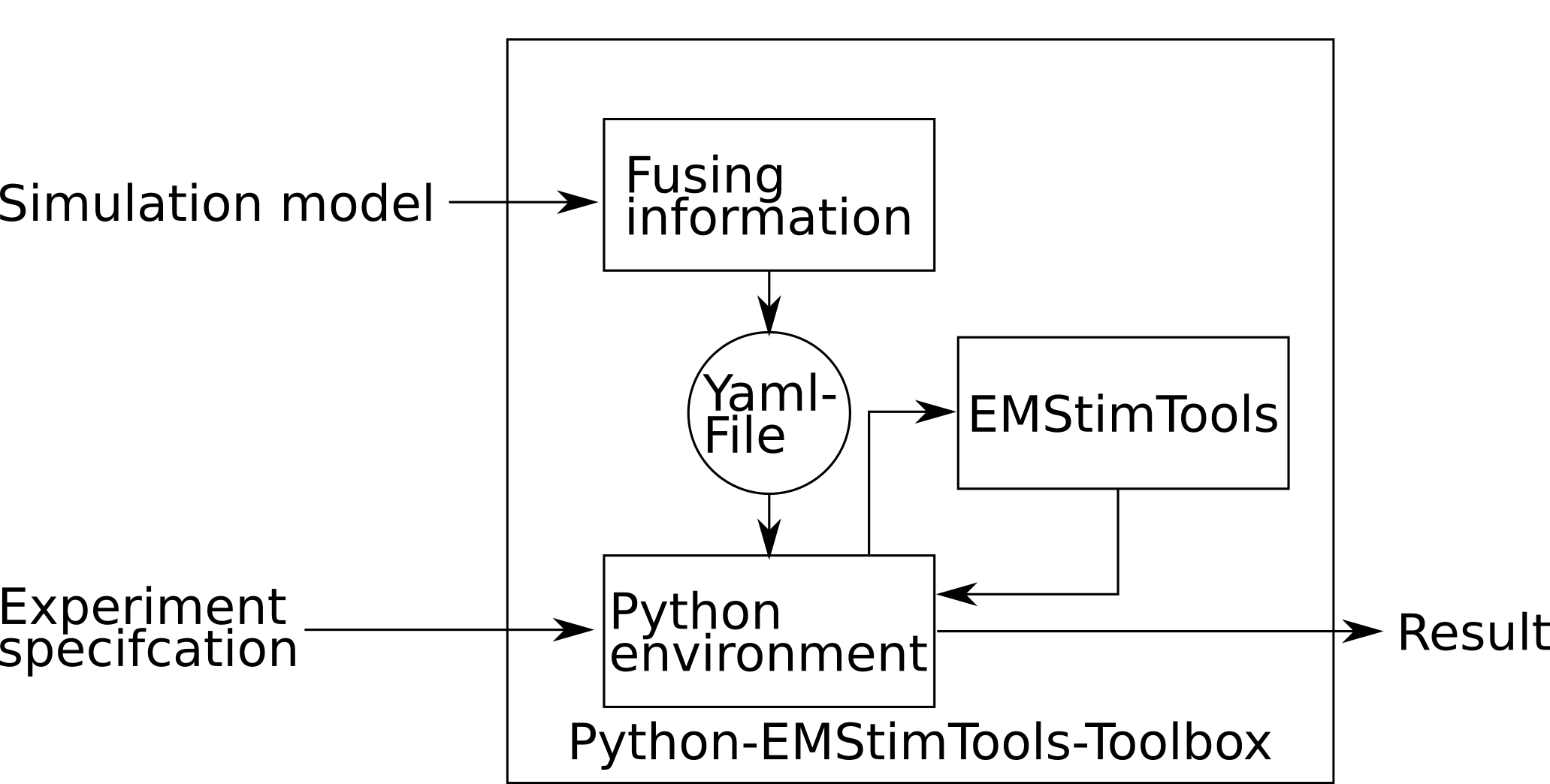}
		\caption{Overview of the structure of Python-EMStimTools-Toolbox.}
		\label{fig:python-emstimtools}
	\end{figure}

	In the simulation experiment artifact, the modeler enters the \stage{Assembling experiment}.
	The milestone of the stage requires that the modeler has to specify the experiment approach and to provide an experiment specification.
	However, to use the simulation experiment for the validation of the simulation model it is additionally required to select the requirement of the simulation model to be checked by the experiment.
	After selecting the requirement, the modeler specifies \enquote{Python-EMStimTools} as the experiment approach.
	For the experiment specification, the modeler writes Python code which calls EMStimTools to run the simulation and compare the results with the analytical solution.
	For this, the \enquote{Python-EMStimTools toolbox} combines the information about the mesh file of the geometrical model, the specification of the physical model, and the specification of the boundary conditions into a single YAML-file.
	This file is afterward passed from the Python environment to the EMStimTools (see Figure~\ref{fig:python-emstimtools}).
	After finishing the specification of the experiment, the modeler leaves the \stage{Assembling experiment}.
	
	Finally, the validation experiment can be executed in the simulation model artifact by entering the \stage{Validating simulation model}.
	However, the validation experiment fails, and therefore the \enquote{failed-milestone} of the stage is achieved.
	The modeler manually refines the mesh by re-entering \stage{Specifying geometrical model} of the geometrical model artifact.
	Thereafter, the validation experiment is run again and now succeeds.
	This time, the stage reaches the \enquote{succeed-milestone} to indicate that the simulation model was successfully validated against the analytical solution.
	
	Similarly, another simulation experiment is assembled to calculate the EF of the ES chamber, i.e., the modeler specifies Python code for collecting the simulation results from EMStimTools and to plot the results via ParaView.
	However, the role of this experiment is simply to analyze the simulation model, i.e., to predict the electric field. Therefore, in contrast to validation or calibration experiments, no requirement on the model output needs to be checked.
	
	\subsection{Exploiting the workflow}
	So far, we used the artifact-based workflow as a mere description of the activities to develop the different products and the information about them.
	However, in practice, we are interested in actually supporting the modeler in conducting the simulation study, e.g., by preventing mistakes leading to invalid products or by reducing the overall effort required for a simulation study.
	In the following, we describe three different methods that exploit the artifact-based workflow to support the modeler in a simulation study.
	
	\paragraph{Guiding the Modeler}
	Probably the most obvious method is to guide the modeler through the simulation study according to the lifecycle models of the artifacts as defined by the workflow.
	The approach ensures that a simulation study is conducted according to the rules as they are expressed in terms of artifacts, stages, guards, and milestones. 
	Thus, a systematic conduction of simulation studies is supported, by which the diverse artifacts that contribute to a simulation study as well as their relation are made explicit \cite{Fujimoto2017}
	
	Technically, we exploit the guards to prevent the modeler from entering inactive stages, i.e., a stage for which the guard is not satisfied, and to derive the missing steps required to enter the stage.
	To prevent the modeler from entering inactive stages, we evaluate the guards of stages against the current state of the simulation study stored in the information models of the artifacts. 
	Only stages of satisfied guards can be entered. 
	Consequently, the guards have to be re-evaluated whenever a milestone has been reached to ensure that the active stages are correctly determined.
	
	However, so far we only constrained the activities of the modeler. 
	We can also use guards, stages, and milestones to infer the missing steps for the modeler to activate a stage.
	Therefore, we perceive the current state of the simulation study as an initial state and the user-desired stage as a goal of a planning problem \cite{hendler1990ai}.
	For the corresponding planning domain, we represent the stages as actions, the guards as their preconditions, and the milestones as their postconditions (see Listing \ref{lst:planning-domain}).
	
	\begin{figure} [h]
		\begin{lstlisting}[caption = Action specification for the \stage{Validating simulation model} in PDDL, label=lst:planning-domain, numbers=left, autogobble=true, frame=single,language=yaml,columns=flexible,aboveskip=3mm, belowskip=3mm,tabsize=2]
		(:action validate-smo
			:parameters (?smo - SModel ?cmo - CModel)
			:precondition 
				(and
					(validated-cmo ?cmo) ; Conceptual model validated?
					(link ?cmo ?smo)
					(assembled-smo ?smo) ; Simulation model assembled?
					(exists (?exp - Exp) ; Is there at least one validation experiment?
						(and
							(link ?smo ?exp) 
							(role ?exp val))))
		:effect (validated-smo ?smo))
		\end{lstlisting}
	\end{figure}
	
	By using a planning system, e.g., the FF planning system \cite{hoffmann2001ff}, we can return the shortest path through the workflow as a suggestion to the modeler.
	Thus, in the above setting at the moment the modeler wants to validate the simulation model, the steps \begin{inparaenum} \item create and assemble a requirement artifact, \item validate the conceptual model, \item assign requirement to the simulation model, and \item assemble a validation experiment, \end{inparaenum} are suggested to the modeler.
	Further, details can be found in~\cite{ruscheinski2019artifact}.
	
	\paragraph{Documentation}
	\label{subsec:documentation}
	During conducting FEA simulation studies, the modeler makes a lot of decisions, some of which are exemplarily captured in our short case study.
	Each of these decisions influences the outcome of the simulation study.
	Thorough documentation of the model building process, as suggested by reporting guidelines for FEA studies~\cite{erdemir2012considerations} facilitates interpreting and assessing the model, its reuse, and enhances the possibilities to reproduce the results.
	
	From the perspective of documentation, our artifact-based workflow describes FEA simulation studies as an abstract process using a set of dos and don'ts of how to conduct a simulation study.
	However, during the simulation study, the process becomes detailed as the modeler moves through a process of concrete activities by which different instances of artifacts are generated, related to each other, and updated. By observing the modeler, the activities form a kind of execution log or audit trail ~\cite{davidson2008provenance} which can be stored to document the simulation study. 
	
	In~\cite{RuscheinskiWilsdorf2019}, we equipped our artifact-based workflow with means to automatically capture this log or trail by observing the modeler moving through the stages of the artifact-based workflow.
	Thereby, we gather provenance information. 
	Provenance subsumes \enquote{information about entities, activities, and people involved in producing a piece of data or thing, which can be used to form assessments about its quality, reliability, or trustworthiness}~\cite{groth2013prov}.
	
	For capturing the information we adopted the PROV-DM formalism~\cite{belhajjame2013prov}.
	According to the PROV-DM formalism, we distinguish between entities, representing physical or digital objects, and activities, representing actions that use entities to produce new ones.
	Entities and activities are related by \emph{Used-Relations}, i.e., an entity was used (entity $\leftarrow $ activity) in an activity, and \emph{WasGeneratedBy-Relations}, i.e., an activity (activity $\leftarrow $ entity) generates an entity. 
	
	To capture the provenance information of a simulation study, we map the artifacts in our workflow to entities and the stages to the activities of the provenance graph.
	Moreover, we identify the inputs (Used-Relation) and the outputs (Generated-Relation) for each stage. 
	Together with transformation rules, this allows us to capture the provenance information of an individual stage. 
	The individual provenance information from the stages is chained together as the modeler moves through the workflow to form the full provenance graph.
	
	The resulting provenance graph becomes quite large over time and thus can be overwhelming for the modeler.
	By exploiting a graph database for storing the information, customized reduction strategies become available to the modeler. 
	On the one hand, the modeler can directly query the provenance graph.
	On the other hand, abstraction-based filtering can easily be applied, e.g., distinguishing sources used for model development, model calibration, and model validation~\cite{RuscheinskiWilsdorf2019}.
	
\begin{figure}
		\includegraphics[width=\textwidth]{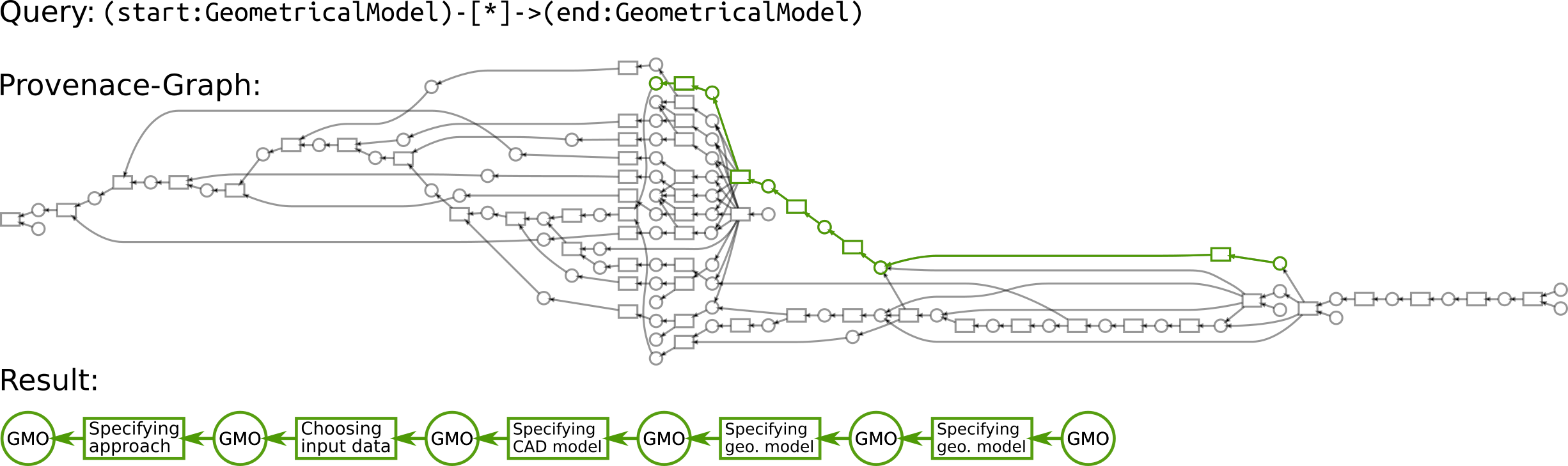}
		\caption{Query-based filtering applied to show the development process across different versions of the geometrical model (GMO).}
		\label{fig:provenance-graph-inputdata}
	\end{figure}

	To apply these methods to the artifact-based workflow for FEA simulation studies only a few changes are needed.
	The transformation rules are extended respectively adapted to cover the new artifacts and stages in the workflow. 
	For example, the \stage{Creating geometrical model}, is translated to an activity with the same name, which uses the entity simulation model and generates two entities, i.e., a new version of the simulation model (which now contains a geometrical model) and a geometrical simulation model. 
	Recording the provenance information of the simplified case study leads to a provenance graph consisting of 130 nodes (75 entities, 55 activities) and 161 edges.
	Query-based filtering can be applied to only show part of the provenance graph corresponding to the source and the specification of the input data  (see Figure~\ref{fig:provenance-graph-inputdata}).
	
	Further, we can automatically extract parts of the model documentation from the provenance graph and the information models of the artifacts.
	For example, we can extract the data used to validate the simulation model or the data used to develop the geometrical model whereas 
	information about the software used to run the simulations can be extracted from the information model of the simulation experiment artifact.
	
	\paragraph{Experiment generation}
	\label{subsec:exp-generation}
	Detailed documentation of simulation studies can be used to support the automatic generation of simulation experiments~\cite{RuscheinskiWSC2018}.
	Therefore, the structure and required inputs need to be identified for specific types of simulation experiments, such as sensitivity analysis or statistical model checking.
	Once identified, they can be represented in the form of software-specific templates, or more abstractly in the form of experiment schemas to use them across different software systems \cite{Wilsdorf2019}.
	Given an experiment template or schema, the information models of the workflow artifacts form a rich source to fill input variables of various experiment schemas.
	The completely filled schemas are used to generate executable experiments in the respective modeling and simulation environment.
	From the perspective of the workflow, there is no difference between the manual and automatic generation of a simulation experiment specifications, as the latter forms simply an alternative to the purely manual experiment specification in the \stage{Specifying simulation experiment} that involves the invocation of an experiment generator.
	A detailed presentation of the pipeline and the generators that lead from composable experiment schemas to executable simulation experiments, such as sensitivity analysis, and which can be applied for stochastic discrete-event simulation as well as FEA studies can be found in \cite{Wilsdorf2019}.
	
	For example, we can generate a convergence test experiment that tests whether further refinement of the mesh leads to a more accurate simulation results \cite{datta2010introduction}.
	Therefore, an adaptive meshing algorithm is used to incrementally refine the mesh until the estimated discretization error lies under a given error threshold or until the maximum number of iterations is reached.
	Moreover, we assume that the entire mesh is refined in every iteration of the algorithm, and the modeler aims at checking the convergence at a given region of interest, e.g., the area marked as \enquote{Contact1}.
	Thus, the corresponding experiment schema requires the following inputs: \begin{inparaenum}
		\item the simulation model, including the physical model and the geometric model,
		\item the region of interest,
		\item an error metric allowing to estimate the discretization error for a given region,
		\item the maximum number of iterations or an error threshold to control 
		\item an initial meshing hypothesis, i.e., the minimal and maximum size of the finite elements, to initialize the meshing algorithm, and
		\item the target experiment specification language
	\end{inparaenum} (see Table \ref{table:conv}).
	
	\begin{table}[!h]
		\caption{Experiment schema defining the inputs for a convergence study.}
		\label{table:conv}
		\begin{tabularx}{\textwidth}{llX}
			\toprule 
			Input         & Type                &  Artifact \\
			\midrule
			Simulation model & Reference & Simulation model \\
			Region of interest & Text & \multirow{3}{*}{Requirement} \\
			Error metric & Math. expression  \\
			Error threshold & Real number  \\
			Max iterations & Real number & -- \\
			Initial max. element size  & Real number &  -- \\
			Initial min. element size  & Real number & -- \\
			Target language & Experiment toolbox & Simulation experiment\\
			\bottomrule
		\end{tabularx}
	\end{table}
	
	In the workflow, some of these input variables can be filled by extracting information from the information model of the simulation experiment artifact, i.e., the corresponding simulation model, the target language chosen by the modeler, and the region of interest, the error metric etc. from the requirement.
	Other inputs might be interpreted as specific for the convergence test, e.g., the number of iterations, or the initial minimum and maximum size of the elements, and thus need to be added manually.
	Once the simulation experiment schema is complete, the experiment generator can be used to generate the experiment specification, e.g. for the EMStimtools (see Code Listing \ref{code:convergence}).

	\begin{figure}[t]
		\begin{lstlisting}[caption = Python code of the automatically generated convergence study (bold: inputs by the modeler)., label=code:convergence, language=Python, numbers=left, frame=single,breaklines=true,autogobble=true,mathescape=false,escapechar=!,	columns=flexible,aboveskip=3mm, belowskip=3mm]
		iterations = !\bfseries 7!
		# initial meshing assumptions
		max_size = !\bfseries 2.4e-2!
		min_size = !\bfseries 1e-3!
		for i in range(iterations):
			# prepare the simulation
			input_dict = prepare_input_dict(max_size, min_size,
				"simulation_model.yml")
			# run the simulation
			model = Simulation(input_dict)
			model.run()
			# get current at Contact1 from results
			current = model.fenics_study.compute_current("Contact1")
			# calculate discretatization error
			if i > 0:
				error = abs(current - current_old)
				results.append((min_size, max_size, current, error))
			current_old = current
			# halve the max_size and min_size for next iteration
			max_size = max_size / 2.0
			min_size = min_size / 2.0
		print(results)
		\end{lstlisting}
	\end{figure}
	
	The above deliberation about where to put which information reveals the artifact-based workflow as a powerful means to explicitly \enquote{model} the process of conducting a simulation structure and thus to structure knowledge about a specific type of simulation study, and to make this knowledge explicit, accessible and reusable.
	
	\section{Related Work}
	\label{sec:related-work}
	In the field of FEA, support has been focused on automating simulation experiments using scientific workflows and scripts  \cite{Reiter2012analyzing, zehner2015workflows}.
	To those simulation experiments belong traditional types of simulation experiments such as sensitivity analysis or optimization \cite{stipetic2015optimization}, but also FEA specific experiment types to assess questions of mesh quality, energy balance, or simulation accuracy~\cite{erdemir2012considerations,burkhart2013finite}, as well as more recent developments, e.g., uncertainty quantification based on surrogate models~\cite{schmidt2012influence}.
	Therefore, the approaches provide the means to specify sub-tasks, e.g., mesh generation \cite{owen1998survey}, and arrange them as a dataflow process network \cite{ludascher2006scientific, lee1995dataflow}, i.e.,  processing pipelines, which we can execute but also adapt and reuse for similar simulation experiments \cite{timmons2017end,scheidegger2008querying}.
	
	Scientific-workflows, such as Taverna \cite{oinn2004taverna} or Kepler \cite{ludascher2006scientific}, provide specialized languages or visual interfaces to describe the processing pipelines and provide additional services, e.g., for distributing the execution of the pipeline \cite{deelman2005pegasus,ludascher2006scientific} or tracking different versions of the workflow specification \cite{barga2010provenance, belhajjame2008metadata}.
	In contrast, scripting languages, such as Python \cite{oliphant2007python} or R \cite{ihaka1996r}, provide a full programming language to describe the arbitrary computational process.
	Consequently, the user can not rely on a closed environment and thus needs to implement the data pipelines with the services and the execution logic.
	
	Another approach to support FEA simulation studies is to use ontologies, which provide a formal specification of the different work products in terms of their attributes and relations.
	Thus, we can use ontologies to share knowledge about the geometrical and physical model across different software suites \cite{TESSIER201376,sun2009framework,Physicsbasedsimulationontology,freitas2013towards}, to derive processing parameters \cite{KESTEL2019292,Capturingsimulationintent} and to document these studies \cite{PENG2017314,mehdi2015linked}. 
	
	Our artifact-based workflow combines and extends existing workflow-based and ontology-based approaches by describing how they are conducted and integrating knowledge about the research question, assumptions, requirements, or input data into this process.
	Therefore, the lifecycle models of our artifacts describe computational processes, as well as interactive processes.
	Existing workflow-approaches can be integrated into the artifact-based workflow to specify and execute simulation experiments.
	Further, the meta-information and the information about the artifacts are made explicit by the information models of the artifacts. 
	Process knowledge about FEA studies and meta-information together allows us to extend existing ontology-based support mechanisms by 
	automatically documenting crucial provenance information and by 
	(semi-)automatically generating various simulation experiments for the FEA study.
	
	\section{Conclusion}
	We adapted the artifact-based workflow developed in~\cite{ruscheinski2019artifact} for FEA simulation studies and showed based on a small case study on FEA of an electric stimulation chamber how such a declarative workflow approach can contribute to the guidance, best practices, and documentation of FEA simulation studies.
	
	To assist in conducting FEA studies, the artifact-based workflow relies on the definition of toolboxes, that allow integrating methods to specify simulation models, provide execution means for analysis, and offer and support diverse experiments, into the workflow.
	Instead of a recipe enlisting a sequence of actions to follow, the artifact-based workflow constrains the execution of activities via guards and milestones.
	Pursuing a least commitment strategy opens up the needed freedom to navigate a FEA study.
	The design, i.e., the artifacts and their relations, ensures that the conceptual model, in terms of objective, inputs, and requirements, as well as simulation experiments, are specified.
	Within the workflow, we define and explicate relations between conceptual model, requirement, inputs, simulation model, simulation experiments, and simulation data.
	
	The number and variety of simulation experiments that are conducted with a simulation model are what determine the understanding of and trust in a simulation model~\cite{anderson2007verification,hicks2015my,burkhart2013finite}.
	Simulation experiments are treated as first-class citizens in conducting a FEA study.
	Based on the defined requirements, their types, and methods that are supported by the experiment toolboxes, the artifact-based workflow approach allows to (semi-)automatically derive, generate, and execute simulation experiments.
	Thus, the conceptual model, in particular, the specified requirements can be used to define (and in the ideal case to generate automatically) a minimal set of simulation experiments to be conducted.
	
	As noted e.g. in~\cite{erdemir2012considerations}, for interpreting and reusing simulation models and results, the thorough documentation of the FEA study is central.
	Mapping artifacts to entities and stages to activities allowed us together with a few transformation rules to automatically store detailed provenance information about FEA studies.
	Thereby, the process by which the simulation model was generated and the activities and resources that contributed to its generation can be traced and queried.
	
	Only part of the parameters that are suggested in~\cite{erdemir2012considerations} for a thorough report of FEA studies is used in our adaptation of the artifact-based workflow for FEA studies.
	Further could be easily integrated into the conceptual model artifact.
	To do so in a structured manner, community efforts are required.
	The development of more standardized vocabularies and ontologies for FEA simulation studies would enable additional inference mechanism, e.g., to automatically check the consistency of the information in the conceptual model, or can serve as a foundation to further develop experiment schemas to generate simulation experiments automatically.
	Therefore, to structure and more formally define context information and thus make it accessible for automatic interpretation is key for effective exploitation of workflow methods for guidance, best practice, and documentation of FEA studies.

	\section*{Acknowledgment}
	
	This work was funded by Deutsche Forschungsgemeinschaft (DFG, German Research Foundation) – SFB 1270/1 - 299150580 and the DFG research project UH 66/18 GrEASE.
	
	\bibliography{Ruscheinski_etal_FEA_Workflow}
\end{document}